\documentclass[aps,pre,reprint,groupedaddress,amsmath,amssymb,showpacs]{revtex4-1}
\usepackage{graphicx}
\usepackage{bm}

\newcommand{\binomial}[2]{
\begin{pmatrix}
#1\\
#2
\end{pmatrix}
}

\begin{document}


\title{Two-lane traffic-flow model with an exact steady-state solution}


\author{Masahiro Kanai}
\email[]{kanai@ms.u-tokyo.ac.jp}
\affiliation{Graduate School of Mathematical Sciences, The University of Tokyo, Komaba 3-8-1, Meguro-ku, Tokyo, Japan}


\date{\today}

\begin{abstract}
We propose a stochastic cellular-automaton model for two-lane traffic flow based on the misanthrope process in one dimension.
The misanthrope process is a stochastic process allowing for an exact steady-state solution; hence we have an exact flow-density diagram for two lane traffic.
In addition, we introduce two parameters that indicate respectively driver's driving-lane preference and passing-lane priority.
Due to the additional parameters, the model shows a deviation of the density ratio for driving-lane use and a biased lane-efficiency in flow.
Then, a mean-field approach explicitly describes the asymmetric flow by the hop rates, the driving-lane preference, and the passing-lane priority.
Meanwhile, the simulation results are in good agreement with an observational data, and we thus estimate these parameters.
We conclude that the proposed model successfully produces two-lane traffic flow particularly with the driving-lane preference and the passing-lane priority.
\end{abstract}


\maketitle
\section{Introduction}
Studies of traffic flow have made much progress so far mainly investigating highway-like single-lane traffic with a wide variety of theoretical and/or practical models \cite{Chowdhury:PR329,Helbing:RMP73,Nagatani:RPP65,Kerner:2004,Kanai:PRE72,Sugiyama:NJP08,Kanai:PRE79}.
These models have been then extended so as to mimic a more realistic traffic flow, in consideration of on-/off- ramps, traffic light control, and multilane traffic \cite{Chowdhury:PR329,Helbing:RMP73,Kerner:2004,Nishi:PRE79}.
In particular, two-lane traffic flow \cite{Cremer:MCS28,Wagner:PA234,Chowdhury:PA235,Nagel:PRE58,Helbing:N396,Fukui:PA303,Knospe:JPA35,Tadaki:JPSJ71}, in which lane change becomes possible, is a typical extension and is quite meaningful in studies of traffic flow; the driver has a complex decision-making process in view of the current traffic situation when attempting lane change, and by which the vehicular motion is distinguished from the substantial motion more clearly than in the case of single-lane traffic.

In most of previous works, it is a common strategy that one includes two general steps in the course of updating: the first step is attempt to change the lane, and the second is forward movement along the lane.
A lane change is accomplished by crossing just sideways through ``wish and decision'' or when safety and incentive criteria are both simultaneously fulfilled;
meanwhile, the forward movement is governed by the same dynamics as a single-lane model.

In order to define a safety criterion in multilane traffic, they assume that the driver takes into account the distances which the driver will have after changing the lane both to the vehicle ahead and behind.
In some models, one considers the relative velocities instead of or in addition to the distances.
Anyway, the safety criterion comes from the first principle of traffic flow dynamics, i.e., {\it the vehicle avoids a collision}, and is consequently regardless of from the right to the left or otherwise.

The incentive criterion reflects the second principle that the driver wishes to increase the velocity up to the legal limit.
Since lane change is possible in two-lane traffic, we can say more precisely that {\it the driver wishes to avoid decreasing the velocity}.
This criterion is also regardless of which lane the vehicle is on.

In addition to the above criteria, we should take legal restrictions into account.
Particularly, a constraint on lane usage, which is often included in the incentive criterion, usually differs in each country as seen below~\cite{Nagel:PRE58}:
(a)In Germany, passing is banned in the right lane.
As a result, they have to change from the right lane to the left not only when a relatively slow vehicle is ahead in the same lane but also when it is in the left lane.
(b)In the United States, by contrast, passing in the right lane is not explicitly forbidden.
We should then consider that the incentive criterion is symmetric to the lanes in the United States.
(c)In Japan (where they drive on the left-hand side), the left lane is assigned by law to be a driving lane; the right lane is, in principle, for passing.
Hence, slow vehicles, e.g. trucks (cf. (a)), are seldom on the passing lane while the density of vehicles is small.

Asymmetric rules on lane usage give rise to a different traffic flow in each lane, and one observes uneven lane use.
The ratio of driving-lane use is large during a small number of vehicles; however it decreases quickly as the vehicle number increases, and falls below that of the passing-lane use in an intermediate density region.
A typical inversion of the ratio for lane use is observed in {\it Autobahn} in Germany~\cite{Nagel:PRE58}.
Also, we see an inversion in terms of flow as well as the density ratio in a Japanese expressway~\cite{Tadaki:JPSJ71}.
(As for the density ratio, one sees rather different behaviors each in Germany and Japan according to the constraint on lane usage.)
Accordingly, it is a validation of two-lane models to reproduce an inversion of the density/flow ratio for lane use.

One of the most successful models for single-lane traffic flow is the Nagel-Schreckenberg (N-S) model~\cite{Nagel:JPI2}, a stochastic cellular-automaton (CA) model.
Since it is elementary as well as suitable for simulations, the N-S model (or its variants) is often incorporated as a forward-motion engine~\cite{Nagel:PRE58,Knospe:JPA35}.
Nevertheless, in comparison with the single-lane case, one has to pay an extra cost of renumbering all the vehicles at each time step.

In this paper, we consider a system of identical particles hopping with nearest-neighbor interactions, focusing on lane-change dynamics.
A complex process to try to change lanes is included in the hope rates which depend exclusively on the particle's neighboring sites.
Consequently, we can specify some parameters characteristic to two-lane traffic flow.
In Sec. II, we define the model.
In Sec. III, we show the simulation results of the model.
Analytical results corresponding to the above simulations are given in Sec. IV, where we consider the mean-field approximation as well as an exact solution.
The paper concludes in Sec. V with some remarks.

\section{Model}

We propose a CA model for two-lane traffic flow based on {\it processus des misanthropes}~\cite{Cocozza-Thivent:ZWG70}: a system of identical particles hopping, on a lattice of finite dimension, with the hop rates depending not only on the occupancy of the departure site but also on that of the target site.

Consider $N$ indistinguishable particles on a one-dimensional periodic lattice with $L$ sites.
Each site contains at most two particles, and one site chosen randomly is updated at each discrete time step, i.e., the random sequential updating.
The particles hop from left to right with a given hop rates $u(n_l,n_{l+1})$, i.e., during the infinitesimal interval $dt$ of time, a particle hops out of site $l$ occupied by $n_l$ particles into site $l+1$ occupied by $n_{l+1}$ particles with probability $u(n_l,n_{l+1})dt$.
(Accordingly, the above hopping is denied with probability $1-u(n_l,n_{l+1})dt$.)
The hop rates should take the value of zero except for $u(1,0)$, $u(1,1)$, $u(2,0)$, and $u(2,1)$.
In Fig. 1(a), we illustrate a misanthrope process with hop rate function $u(n_l,n_{l+1})$.
This configuration is denoted by $(n_l)=(0,1,2,1,1,0,2,2,1)$.

\begin{figure}
\includegraphics[scale=0.4]{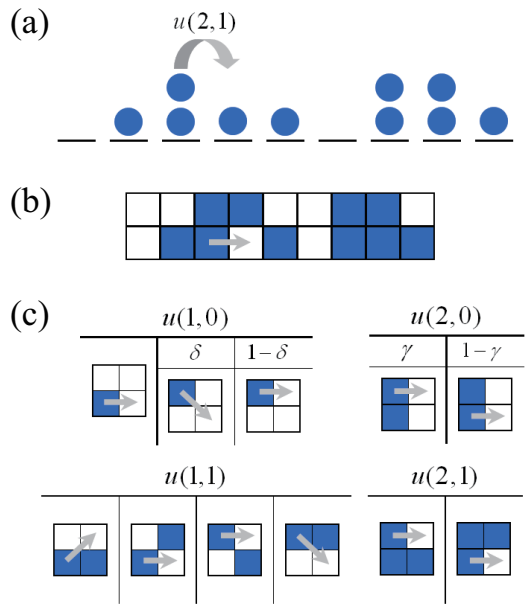}
\caption{
(Color online) (a)A configuration of particles in the misanthrope process with 9 sites and 10 particles.
One particle is now hopping from a site occupied by two to the next site occupied by one with rate $u(2,1)$.
(b)A configuration of vehicles in the present model for two-lane traffic flow.
We consider the lower cell is the driving lane and the upper cell is the passing lane at each site.
Colored cells are occupied by a vehicle; white cells are empty.
The model coincides with the above misanthrope process unless one regards lanes.
(c)Vehicular motions and the hop rates in the present model.
(Any other motion is prohibited.)
If there are two possible motions, the hop rate is modified with a division parameter, $\delta$ or $\gamma$.
\label{fig1}}
\end{figure}

We do not yet take into account which lane each particle is on and whether they change lanes or not; only how many particles are contained in each road section, i.e., at each site of the lattice.
In Fig. 1(b), we describe a configuration in the two-lane road which corresponds to Fig. 1(a), and now the lane number is assigned to each particle.
In the following part, particles with the lane number (Fig. 1(b)) are referred to as vehicle in distinction from those in the misanthrope process (Fig. 1(a)).
Provided that a vehicle does not cut in front of another one, we can specify the vehicle to hop at each time step, except for the configurations with the target site empty: $(1,0)$ and $(2,0)$.
Accordingly, we see that the model admits two additional parameters, and introduce {\it driving-lane preference} (DLP) $\delta$ and {\it passing-lane priority} (PLP) $\gamma$.
The DLP means the probability of a particle on the passing lane to attempt a lane change to the driving lane; meanwhile, the PLP presents a priority level of the passing lane.
In Fig. \ref{fig1}(c), we summarize all possible motions of vehicles in the model and the rates to actually move.
It is remarkable that the present model is established without any conditional.

Provided a constraint on the hop rates,
\begin{equation}
u(2,1)=u(2,0)-u(1,0),
\label{condition}
\end{equation}
the misanthrope process has an invariant product measure, or in other words, an exact steady-state solution of factorized form~\cite{Cocozza-Thivent:ZWG70}.
Accordingly, as far as the mean values in two lanes (e.g., the total flow) are concerned, we can obtain exact solutions under the condition (\ref{condition}).
The solvability condition (\ref{condition}) is derived purely through a mathematical argument.
However, it is acceptable for the traffic model because generally speaking, drivers do not like to drive side by side, which suggests $u(2,0)>u(1,0)$.
In addition, since it cannot be comfortable to drive while surrounded by other vehicles, we can presume $u(1,0)>u(2,1)$.
In the following part of this paper, we assume that the hop rates satisfy (\ref{condition}), giving a priority to the exact solution.

\section{Simulation results}

\begin{figure*}[ht]
\includegraphics[scale=0.8]{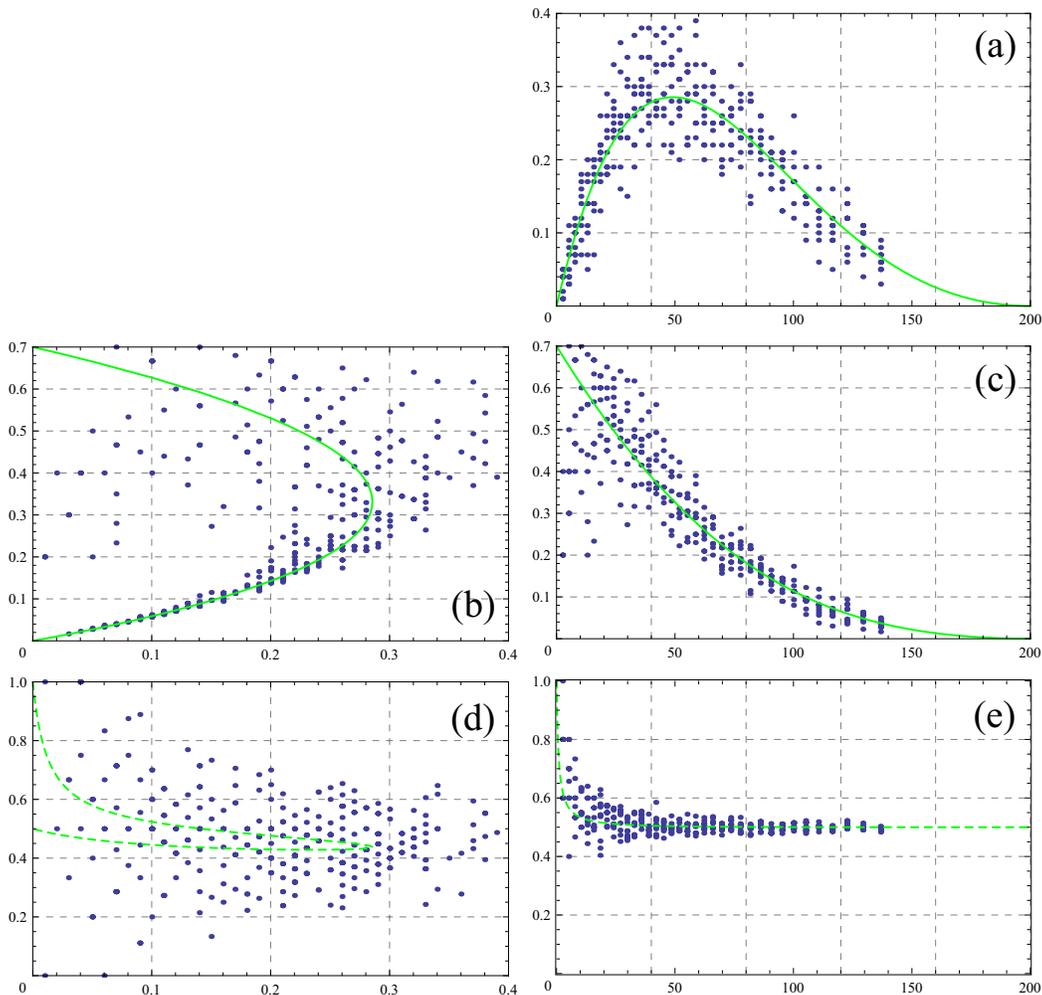}
\caption{
(Color online) The simulation results (dot) of the proposed model for two-lane traffic with $L=100$, $u(2,0)=1.$, $u(1,0)=0.6$, $u(1,1)=0.7$, $u(2,1)=0.4$, $\delta=0.02$, and $\gamma=0.75$: (a)flow vs density, (b)velocity vs flow, (c)velocity vs density, (d)flow ratio for driving-lane use, and (e)density ratio for driving-lane use.
The number of vehicles takes the values $0\leq N\leq 180$, changing by $5$.
We use $9$ random configurations as the initial one.
After $500$ time steps, the data are collected; in particular, flow is measured as the average number of hops over $L$ time steps.
Also, we give analytical results corresponding to the above simulations: exact solutions (solid line) in (a), (b), and (c), and mean-field calculations (dashed line) in (d) and (e).
\label{fig2}}
\end{figure*}

In Fig. 2, we show the simulation results of the present model: (a)flow vs density, (b)average velocity vs flow, (c)velocity vs density, (d)flow ratio for driving-lane use, and (e)density ratio for driving-lane use.
Both flow and velocity are for two lanes.

It is important to note that in Fig. 2, we plot the results with a calibration of the density \cite{Kanai:JPSJ2010}.
The calibration method was developed to simulate particle flows such as vehicular traffic truly with a simple CA model.
The cell length (i.e., site size) was changed depending on the vehicle density, and thus the density $\rho_{\mbox{\scriptsize{CA}}}=N/L$ used in the simulation is calibrated as
\begin{equation}
\rho_{\scriptsize \mbox{RW}}=2\Bigl(1-\sqrt{1-\frac{\rho_{\mbox{\scriptsize CA}}}2}\Bigr),
\label{calibration}
\end{equation}
where $\rho_{\scriptsize \mbox{RW}}$ means the {\it real-world} density, which is used in the above plots.
In \cite{Kanai:JPSJ2010}, the calibration for single-lane models is given by $\rho_{\scriptsize \mbox{RW}}=1-\sqrt{1-\rho_{\mbox{\scriptsize CA}}}$.
We hence replace the single-lane densities $\rho_{\scriptsize \mbox{RW}}$ and $\rho_{\mbox{\scriptsize CA}}$ with $2\rho_{\scriptsize \mbox{RW}}$ and $2\rho_{\mbox{\scriptsize CA}}$ in order to have a calibration for two-lane models.

One of our goals is to compare the simulation results with the observational data given in \cite{Wiedemann:1995,Nagel:PRE58}, which present the characteristics of two-lane traffic flow as described in the Introduction.
We consequently find that the model proposed reproduces a typical two-lane traffic flow.
In particular, the inversion of the flow ratio for driving-lane use (Fig. 2(d)) shows an excellent agreement with the observational data.
In contrast, one does not see such a sharp inversion of the density (Fig. 2(e)).
Nevertheless, we find a wide fluctuation of the ratio in a low density region, which also occurs in the observational data.
It implies that the DLP and PLP surely make a symmetric flow unstable.

Moreover, we consider that a strong inversion of the density ratio observed in real-world traffic can be attributed largely to two factors we do not take in: one is the existence of (relatively) slow vehicles, and the other is a restriction on their lane usage.
In a low density region, since slow vehicles cannot catch up with fast ones, slow ones hardly have an interaction with fast ones; by contrast, fast ones are always under the influence of slow ones.
As a result, slow vehicles tend to get in the driving lane; meanwhile, fast vehicles remain in the passing lane.

\section{Analytical results}

\subsection{Exact solution: the flow-density plot}

As long as the hop rates satisfy the condition (\ref{condition}), we can obtain an exact solution of the present model.
We mean that from the master equation for the model, the steady-state probability $P(\omega)$ of finding the system in a configuration $\omega=(n_1,n_2,\ldots,n_L)$ is obtained in an explicit form and then we can calculate any expectation value in principle.

In the steady state, the master equation for the misanthrope process including the present model becomes
\begin{equation}
\begin{aligned}
\sum^L_{l=1}\Bigl[u(n_{l-1}+1,n_l-1)P(\ldots,n_{l-1}+1,n_l-1,\ldots)\\
-u(n_{l-1},n_l)P(\ldots,n_{l-1},n_l,\ldots)\Bigr]=0,
\end{aligned}
\label{mastereq}
\end{equation}
where $0<u(m,n)<1$ if $m=1,2$ and $n=0,1$; otherwise $u(m,n)=0$ (including the cases, $m<0$ or $n<0$).
Following \cite{Cocozza-Thivent:ZWG70,Evans:JPA38}, we let
\begin{equation}
P(\omega)\propto \prod^L_{l=1}f(n_l),
\label{ff}
\end{equation}
where $f(n)$ is called the single-site weight and one factor for each site of the system.
Then, one immediately finds a solution satisfying (\ref{mastereq}):
\begin{equation}
\begin{aligned}
f(2)=&\frac{u(1,1)f(1)^2}{u(2,0)f(0)},\\
u(2,1)=&u(2,0)-u(1,0).
\end{aligned}
\label{f2}
\end{equation}
The second equation is identical to (\ref{condition}), i.e., (\ref{condition}) is a necessary condition.
Denote by $\sigma_n(\omega)$ the number of sites containing $n$ particles in a configuration $\omega$.
Since $\sigma_0(\omega)+\sigma_1(\omega)+\sigma_2(\omega)=L$ and $\sigma_1(\omega)+2\sigma_2(\omega)=N$, the first equation in (\ref{f2}) leads to
\begin{equation}
\begin{aligned}
\prod_lf(\omega_l)
=&f(0)^{L-N}f(1)^{N}\Bigl(\frac{u(1,1)}{u(2,0)}\Bigr)^{\sigma_{2}(\omega)}.
\end{aligned}
\end{equation}
Accordingly, we have
\begin{equation}
P(\omega)=\frac{1}{Z_{LN}}\Bigl(\frac{u(1,1)}{u(2,0)}\Bigr)^{\sigma_{2}(\omega)},
\label{P}
\end{equation}
where the normalization is given by
\begin{equation}
Z_{LN}=\sum_{|\omega|=N}\Bigl(\frac{u(1,1)}{u(2,0)}\Bigr)^{\sigma_{2}(\omega)}.
\label{Z}
\end{equation}
The summation in (\ref{Z}) is done over all the configurations $\omega=(n_1,n_2,\ldots,n_L)$ such that $|\omega|=\sum_ln_l=N$.
One often refers to $Z_{LN}$ as the {\it nonequilibrium} partition function, but nevertheless the system is in a steady state but far from equilibrium.

The partition function $Z_{LN}$ is described by the Gauss hypergeometric function ${}_2F_1(\alpha,\beta,\gamma;z)$ as
\begin{equation}
Z_{LN}=\binomial{L}{N}{}_2F_1\Bigl(\frac{-N}{2},\frac{-N+1}{2},L-N+1;4\frac{u(1,1)}{u(2,0)}\Bigr).
\label{ZF}
\end{equation}
Now, we can calculate any expectation value, in principle, using the partition function \cite{Kanai:JPA39,Kanai:JPA40}.
The flow $Q_{LN}$, defined by
\begin{equation}
Q_{LN}=\sum_{|\omega|=N}u(n_1,n_2)P(\omega),
\end{equation}
is obtained as a function of the density $\rho=N/L$ in the limit where $L,N$ tends to infinity:
let $Q(\rho)=\lim_{L\rightarrow\infty}Q_{L,\rho L}$ then we obtain
\begin{equation}
\begin{aligned}
Q(\rho)=&\frac{u(2,0)}2\rho(2-\rho)\\
&\cdot\Biggl(1-\frac{2-\rho-2\frac{u(1,0)}{u(2,0)}(1-\rho)}{1+\sqrt{1-\bigl(1-4\frac{u(1,1)}{u(2,0)}\bigr)\rho(2-\rho)}}\Biggr).
\label{Q}
\end{aligned}
\end{equation}
This gives the flow-density plot.
In Fig. 2, we show the exact solution through the calibration (\ref{calibration}).
Refer to the Appendix \ref{Appendix:flow} for the detail of the calculation in this subsection.

\subsection{Mean-field approximation: lane occupancy and lane efficiency}

We now turn to the calculations on lane occupancy and lane efficiency.
Here the lanes are distinguished, and then it seems hopeless to exactly solve the issue in the same manner as above.
We hence use the mean-field approximation in which all correlations between sites are neglected, i.e., we suppose
\begin{equation}
P(\omega)=\prod^L_{l=1}\pi(\omega_l),
\label{mf}
\end{equation}
where $\pi(\mbox{c})$ is a probability for a site of the lattice to be in configuration $\mbox{c}\in\{\mbox{e},\,\mbox{d},\,\mbox{p},\,\mbox{f}\}$; $\mbox{e}$ [$\mbox{f}$] means that the site is empty [fully occupied], and $\mbox{d}$ [$\mbox{p}$] means that one particle occupies the driving [passing] lane of the site.
(Note that the distribution $\pi$ is common in sites because of the periodic boundary condition.)

Immediately one sees $\pi(\mbox{e})+\pi(\mbox{d})+\pi(\mbox{p})+\pi(\mbox{f})=1$ and $\rho=\pi(\mbox{d})+\pi(\mbox{p})+2\pi(\mbox{f})$.
Then, solving (\ref{mastereq}) under (\ref{mf}) leads to
\begin{gather}
\pi(\mbox{d})=\frac{2\rho}{2\rho+\delta \frac{u(1,0)}{u(1,1)}\Bigl(1-\rho+\sqrt{1+(4\frac{u(1,1)}{u(2,0)}-1)\rho(2-\rho)}\Bigr)},
\end{gather}
and
\begin{gather}
\pi(\mbox{f})=\frac12\Bigl(\rho-\frac{\rho(2-\rho)}{1+\sqrt{1+(4\frac{u(1,1)}{u(2,0)}-1)\rho(2-\rho)}}\Bigr).
\end{gather}
We thus obtain $\pi$ for a fixed $\rho$ and are able to calculate expectation values in the mean-field approximation.

In the approximation, the density ratio $R_{\mbox{\scriptsize d}}$ of driving-lane use is calculated as follows:
\begin{widetext}
\begin{gather}
\begin{aligned}
R_{\mbox{\scriptsize d}}(\rho)=&\frac{\pi(\mbox{d})+\pi(\mbox{f})}{\rho}\\
=&\frac12\Biggl(1+\frac{\delta\frac{u(1,0)}{u(2,0)}(2-\rho)^2}{\bigl(4\frac{u(1,1)}{u(2,0)}-1\bigr)\rho(2-\rho)+\bigl[\delta\frac{u(1,0)}{u(2,0)}(2-\rho)+\rho\bigr]\bigl[1+\sqrt{1+\bigl(4\frac{u(1,1)}{u(2,0)}-1\bigr)\rho(2-\rho)}\bigr]}\Biggr)\\
=&\left\{
\begin{aligned}
&1-\Bigl(\frac1\delta \frac{1}{u(1,0)}+\frac{1}{u(2,0)}\Bigr)u(1,1)\rho+O(\rho)^2\\
&\frac12+\frac\delta8\frac{u(1,0)}{u(2,0)}(2-\rho)^2+O(2-\rho)^3.
\end{aligned}
\right.
\end{aligned}
\label{rd}
\end{gather}
\end{widetext}
In Fig. 2(e), we show the graph of (\ref{rd}) by a dashed line, which agrees with the simulation result.
Then, from (\ref{rd}) we suppose that an inversion of the density ratio does not occur in the present model, although the simulation result shows a large fluctuation of the ratio in a low density region.
(One sees that $R_{\mbox{\scriptsize d}}(\rho)$ is always larger than $1/2$.)
Finally, expansion of $R_{\mbox{\scriptsize d}}$ in $\rho$ reveals how the hop rates included contribute around $\rho=0,2$.
These help us determine the parameter values, especially $\delta$.

As well, the flow ratio $R_{\mbox{\scriptsize f}}$ of driving-lane use, defined by
\begin{widetext}
\begin{gather}
R_{\mbox{\scriptsize f}}(\rho)=\frac{u(1,0)\pi(\mbox{d})\pi(\mbox{e})+u(1,1)\pi(\mbox{d})(\pi(\mbox{d})+\pi(\mbox{p}))+u(2,0)\pi(\mbox{f})\pi(\mbox{e})+u(2,1)\pi(\mbox{f})\pi(\mbox{p})}{Q(\rho)}
\label{rf}
\end{gather}
\end{widetext}
 can be calculated as a function of $\rho$.
In Fig. 2(d), we show the $R_{\mbox{\scriptsize f}}$-$Q$ plot using a parametric plot with $\rho$.
The explicit expression of $R_{\mbox{\scriptsize f}}(\rho)$ is, however, too complicated to read.
We instead give the graph of (\ref{rf}) in Fig. 3 and expansions of $R_{\mbox{\scriptsize f}}(\rho)$ around $\rho=0,2$:
\begin{gather}
R_{\mbox{\scriptsize f}}(\rho)=\left\{
\begin{aligned}
&1-\frac{(1+\gamma\delta)u(1,1)}{\delta u(1,0)}\rho+O(\rho)^2\\
&\frac12-\Biggl(\frac\delta4\frac{u(1,0)}{u(2,0)}+\Bigl(\gamma-\frac12\Bigr)\frac{u(1,1)}{u(2,1)}\Biggr)(2-\rho)\\
&\quad+O(2-\rho)^2.
\end{aligned}
\right.
\label{rf2}
\end{gather}
These also help us with parameter fitting.
Working in the mean-field approximation, we see that the flow ratio has a very large fluctuation around the mean value and that the ratio takes similar values at the same flow but different densities.
\begin{figure}[htb]
\includegraphics[scale=0.6]{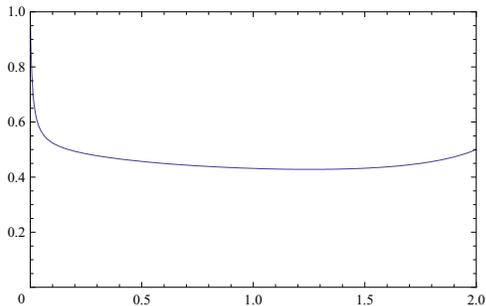}
\caption{
(Color online) Mean-field calculation of the density ratio $R_{\mbox{\scriptsize f}}(\rho)$ for the driving-lane use.
\label{fig3}
}
\end{figure}

\section{Conclusion}

In this paper, we propose a two-lane traffic-flow model with six rate parameters presenting free hop, overtaking, getting out of a side-by-side configuration, passing consecutive vehicles, the DLP, and the PLP, respectively denoted by $u(1,0)$, $u(1,1)$, $u(2,0)$, $u(2,1)$, $\delta$, and $\gamma$.
The model differs substantially from previous ones in the following ways:
(a)In the present model, vehicles are located by site and lane (a so-called {\it site-oriented} description) but not by the vehicle identification number (a {\it car-oriented} description) \cite{Schadschneider:JPA30}.
(b)The vehicles undertake a lane change either when overtaking or when returning to the driving lane voluntarily.
Then, overtaking is certainly done.
Consequently, the DLP contributes to a deviation of the density ratio for driving lane use especially in a low density region.
(c)The PLP, as well as the DLP, explicitly gives rise to an asymmetric flow, while the overtaking parameter does not take the role.
More precisely, the PLP contributes to a deviation of the flow ratio for driving lane use in a high density region.

In the simulation, we optimize the parameter values in order to reproduce an observational data given in \cite{Nagel:PRE58}, and thus find that the DLP is very small and the PLP is relatively large.
Accordingly, we think that the reason for changing lanes is to overtake the slower vehicle in front, rather than to avoid deceleration.
We also take it that drivers have a tacit agreement that vehicles on the passing lane are willing to overtake those on the driving lane.
In fact, drivers will feel unsafe if they drive side by side.

Solving the master equation directly, we obtain an exact solution especially for the flow-density diagram.
If one considers which lane a vehicle is on, the mean-field approach alternatively provides detailed information on how the parameters contribute to an asymmetric flow.
The results obtained may be fairly precise since in fact the approximation happens to show the exact solution (\ref{Q}).
(It may not be so surprising but is never trivial. See, e.g., \cite{Plischke:2006} for further discussion of mean-field theories.)

There are further problems to be addressed:
(a)It is understood that the parallel update rule is better than the others for CA modeling of traffic flow \cite{Schreckenberg:PRE51}.
Accordingly, we should first find an exact solution of the misanthrope process with parallel updating.
(b)An open boundary condition should be considered from both theoretical and practical viewpoints.
(c)Modeling of more than two lane, particularly three-lane, traffic flow will be challenging because one is supposed to deal with a conflict happening when two vehicles attempt to get into the middle lane.
(Then, the shuffled dynamics \cite{Wolki:JPA39}, where vehicles are updated in random order, may be helpful.)

\begin{acknowledgements}

We thank K. Nishinari and R. Nishi for helpful discussions.
This work is supported by Global COE Program, ``The research and training center for new development in mathematics'', at Graduate School of Mathematical Sciences, The University of Tokyo.
\end{acknowledgements}

\appendix*

\section{Exact solution for the flow-density diagram}
\label{Appendix:flow}

In order to solve (\ref{mastereq}) with (\ref{ff}), we introduce a counterterm which cancels under the sum and thus have
\begin{gather}
\begin{aligned}
&\bar{f}(n_{l-1})f(n_l)-f(n_{l-1})\bar{f}(n_l)\\
&\qquad=u(n_{l-1}+1,n_l-1)f(n_{l-1}+1)f(n_l-1)\theta(n_l)\\
&\qquad\qquad\qquad -u(n_{l-1},n_l)f(n_{l-1})f(n_l)\theta(n_{l-1}),
\end{aligned}
\label{fbar}
\end{gather}
having cancelled common factors (a product over the function $f(n_k)$ at all sites $k\ne l-1,l$).
Here, $\bar{f}(n)$ is some auxiliary function to be determined.

Substitution of $u(m,n)=0$ for $m\geq3$ or $n\geq 2$ and $f(n)=0$ for $n\geq 3$ into (\ref{fbar}) yields
\begin{gather}
\begin{aligned}
\bar{f}(1)f(0)-f(1)\bar{f}(0)=&-u(1,0)f(1)f(0),\\
\bar{f}(0)f(2)-f(0)\bar{f}(2)=&u(1,1)f(1)^2,\\
\bar{f}(2)f(0)-f(2)\bar{f}(0)=&-u(2,0)f(2)f(0),\\
\bar{f}(1)f(2)-f(1)\bar{f}(2)=&u(2,1)f(2)f(1).
\end{aligned}
\end{gather}
As a result, we obtain (\ref{f2}) and
\begin{gather}
\begin{aligned}
\bar{f}(2)=&\Bigl(\frac{\bar{f}(0)}{f(0)}-u(2,0)\Bigr)\frac{u(1,1)f(1)^2}{u(2,0)f(0)},\\
\bar{f}(1)=&\Bigl(\frac{\bar{f}(0)}{f(0)}-u(1,0)\Bigr)f(1).
\end{aligned}
\end{gather}

Let us define $\lambda=u(1,1)/u(2,0)$ in the following calculation.
Consider $d$ sites are fully occupied in configuration $\omega$, i.e., $d=\sigma_2(\omega)$ and $0\leq d\leq \lfloor N/2\rfloor$, then, there are $\scriptsize\binomial{L}{d}\binomial{L-d}{N-2d}$ corresponding configurations.
The partition function given in (\ref{Z}) is hence calculated as
\begin{equation}
\begin{aligned}
Z_{LN}=&\sum^{\lfloor N/2\rfloor}_{d=0}\binomial{L}{d}\binomial{L-d}{N-2d}\lambda^{d}\\
=&\binomial{L}{N}\sum^{\lfloor N/2\rfloor}_{d=0}\frac{(-N)_{2d}}{(L-N+1)_d}\frac{\lambda^{d}}{d!},
\end{aligned}
\end{equation}
where the Pochhammer symbol $(a)_n=a(a+1)\cdots(a+n-1)$.
Using the identity
\begin{equation}
(a)_{2n}=2^{2n}\Bigl(\frac{a}{2}\Bigr)_n\Bigl(\frac{a+1}{2}\Bigr)_n,
\end{equation}
one finds (\ref{ZF}).
(Note that the Gauss hypergeometric function is defined by ${}_2F_1(\alpha,\beta,\gamma;x)=\sum^\infty_{n=0}\frac{(\alpha)_n(\beta)_n}{(\gamma)_n}\frac{x^n}{n!}$.)

The flow for the system with given $L$ and $N$ is presented by the partition function as
\begin{equation}
\begin{aligned}
Q_{LN}=&\sum_{|\omega|=N}u(\omega_1,\omega_2)P(\omega)\\
=&\sum^N_{m,n=0}u(m,n)f(m)f(n)\frac{Z_{L-2,N-m-n}}{Z_{LN}}\\\\
=&u(1,0)\frac{Z_{L-2,N-1}-\lambda Z_{L-2,N-3}}{Z_{LN}}\\
&\qquad +u(2,0)\frac{2\lambda Z_{L-2,N-2}+\lambda Z_{L-2,N-3}}{Z_{LN}}.
\end{aligned}
\end{equation}
This is the most general form, but however, the expression is not clear to understand.

There is a useful identity of the Gauss hypergeometric function:
\begin{gather}
\begin{aligned}
&(1+x)^\alpha {}_2F_1(\alpha,\alpha-\gamma+1,\gamma;x)\\
&\qquad\qquad\qquad ={}_2F_1\Bigl(\frac{\alpha}2,\frac{\alpha+1}2,\gamma;\frac{4x}{(1+x)^2}\Bigr),
\end{aligned}
\end{gather}
by which we transform (\ref{ZF}) into a simple one,
\begin{gather}
Z_{LN}=\binomial{L}{N}(1+z)^{-N} {}_2F_1(-N,-L,L-N+1;z),
\end{gather}
where
\begin{equation}
\lambda=\frac{z}{(1+z)^2}.
\end{equation}
Then, using the following identities:
\begin{gather}
\begin{aligned}
&\gamma(\gamma-1)[{}_2F_1(\alpha,\beta,\gamma-1;x)-{}_2F_1(\alpha,\beta,\gamma;x)]\\
&\qquad\qquad =\alpha\beta x {}_2F_1(\alpha+1,\beta+1,\gamma+1;x),
\end{aligned}\\
\begin{aligned}
&(\alpha-\beta)(1-x){}_2F_1(\alpha,\beta,\gamma;x)+(\gamma-\alpha){}_2F_1(\alpha-1,\beta,\gamma;x)\\
&\qquad\qquad +(\beta-\gamma){}_2F_1(\alpha,\beta-1,\gamma;x)=0,
\end{aligned}\\
\begin{aligned}
x{}_2F_1'(\alpha,\beta,\gamma;x)=\alpha[{}_2F_1(\alpha+1,\beta,\gamma;x)-{}_2F_1(\alpha,\beta,\gamma;x)],
\end{aligned}
\end{gather}
we finally obtain the exact flow for arbitrary $L$ and $N$,
\begin{equation}
\begin{aligned}
Q_{LN}=&u(1,0)\frac{N(L-N)}{L(L-1)}\frac{1+z}{1-z}\Bigl(1-\frac{2z}{N}\frac{F'}{F}\Bigr)\\
&\quad -u(2,0)\frac{N(2L-N)}{L(L-1)}\frac{z}{1-z}\Bigl(1-\frac{1+z}{N}\frac{F'}{F}\Bigr),
\end{aligned}
\label{QLN}
\end{equation}
where $F={}_2F_1(-N,-L,L-N+1;z)$ and $F^\prime$ denotes the derivative of $F$ with respect to $z$.

In order to find the flux in the thermodynamic limit where $L$ and $N$ tend to infinity with fixed $\rho=N/L$, we start with the Gauss hypergeometric differential equation for $F$,
\begin{gather}
\begin{aligned}
z(1-z)F''+[1+L-N-(1-L-N)z]F'\\
-LN F=0,
\label{HGDE}
\end{aligned}
\end{gather}
to consider $g=F'/F$ appearing in (\ref{QLN}), which is expected to remain finite in the thermodynamic limit if divided by $N$.
One sees from (\ref{HGDE}) 
\begin{gather}
\begin{aligned}
z(1-z)(g'+g^2)+[1+L-N-(1-L-N)z]g\\
-LN=0.
\end{aligned}
\end{gather}
Hence we let $g=g_{1}N+g_0+g_{-1}N^{-1}+g_{-2}N^{-2}+\cdots$, finding the following quadratic equation for $g_1$ in the $N^1$ order in the thermodynamic limit:
\begin{equation}
\rho z(1-z){g_1}^2+[1-\rho+(\rho+1)z]g_1-1=0.
\label{g1}
\end{equation}
Solving (\ref{g1}) gives
\begin{equation}
g_1=\frac{\rho-\frac{1+z}{1-z}+\sqrt{(\frac{1+z}{1-z})^2-2\rho+\rho^2}}{2\rho z}.
\end{equation}
Now we are ready for taking the thermodynamic limit of (\ref{QLN}).
The result is in (\ref{Q}).

\end{document}